%% file: B2G-12-014_temp.tex
\pdfoutput=1

\documentclass[11pt,twoside,a4paper,cmspaper,final,collab]{cms-tdr}
\def\svnVersion{249989}\def\cmsCernNo{2013-206}\def\cmsCernDate{\today}\def\cmsMessage{Published in the Journal of High Energy Physics as \href{http://dx.doi.org/10.1007/JHEP06(2014)125}{\doi{10.1007/JHEP06(2014)125.}}}

\begin{document}\cmsNoteHeader{B2G-12-014}

\hyphenation{had-ron-i-za-tion}
\hyphenation{cal-or-i-me-ter}
\hyphenation{de-vices}

\RCS$Revision: 217356 $
\RCS$HeadURL: svn+ssh://svn.cern.ch/reps/tdr2/papers/B2G-12-014/trunk/B2G-12-014.tex $
\RCS$Id: B2G-12-014.tex 217356 2013-11-20 16:18:27Z alverson $
\newcommand{\cPqtstar}{\ensuremath{\cmsSymbolFace{t}^*}} 
\newcommand{\cPaqtstar}{\ensuremath{\overline{\cmsSymbolFace{t}}{}^*}} 
\providecommand{\ttbarstar}{\ensuremath{\cPqt^*\cPaqt^*}\xspace}
\providecommand{\re}{\ensuremath{\cmsSymbolFace{e}}}
\cmsNoteHeader{B2G-12-014} 
\title{Search for pair production of excited top quarks in the lepton+jets final state}

\author*[ntu]{Kai-Feng Chen}
\author*[ntu]{Ulysses Grundler}
\author*[ntu]{Yeng-Ming Tzeng}
\author*[ntu]{Ta-Wei Wang}

\date{\today}

\abstract{
A search is performed for pair-produced spin-3/2 excited top
quarks~(\ttbarstar),
each decaying to a top~quark and a gluon.
The search uses data collected with the CMS detector from pp collisions
at a center-of-mass energy of $\sqrt{s} = 8$\TeV, selecting events that have a single isolated
muon or electron, an imbalance in transverse momentum,
and at least six jets, of which one must be compatible with
originating from the fragmentation of a b quark.
The data, corresponding to an integrated luminosity of 19.5\fbinv,
show no significant excess over standard model predictions,
and provide a lower limit of 803\GeV at 95\% confidence on the
mass of the spin-3/2 \cPqtstar~quark
in an extension of the Randall--Sundrum model,
assuming a 100\%
branching fraction of its
decay into a top quark and a gluon.
This is the first search for a spin-3/2 excited top quark performed at the LHC.
}

\hypersetup{%
pdfauthor={CMS Collaboration},%
pdftitle={Search for pair production of excited top quarks in the lepton+jets final state},%
pdfsubject={CMS},%
pdfkeywords={CMS, physics, beyond two generations, excited top quark}}

\maketitle 

\input{variables}

\input{introduction}
\input{data}
\input{selection}

\input{reconstruction}
\input{background}
\input{uncertainties}

\input{analysis}
\input{summary}

\bibliography{auto_generated}   

\cleardoublepage \appendix\section{The CMS Collaboration \label{app:collab}}\begin{sloppypar}\hyphenpenalty=5000\widowpenalty=500\clubpenalty=5000\input{B2G-12-014-authorlist.tex}\end{sloppypar}
\end{document}

%% file: variables.tex
\newcommand{\muLumi}[1][0]{\ensuremath{19.5\ifthenelse{#1=0}{\fbinv}{}}}
\newcommand{\elLumi}[1][0]{\ensuremath{19.5\ifthenelse{#1=0}{\fbinv}{}}}

\newcommand{\muLumiUnc}{\ensuremath{0.5}}
\newcommand{\elLumiUnc}{\ensuremath{0.5}}

\newcommand{\muYield}{13\,636}
\newcommand{\elYield}{11\,643}

\newcommand{\muYieldPred}{$15\,100\pm4\,400$}
\newcommand{\elYieldPred}{$13\,100\pm3\,700$}

\newcommand{\muLimitObs}[1][0]{\ensuremath{680\ifthenelse{#1=0}{\GeV}{}}}
\newcommand{\elLimitObs}[1][0]{\ensuremath{749\ifthenelse{#1=0}{\GeV}{}}}
\newcommand{\combLimitObs}[1][0]{\ensuremath{803\ifthenelse{#1=0}{\GeV}{}}}

\newcommand{\muLimitExp}[1][0]{\ensuremath{689\ifthenelse{#1=0}{\GeV}{}}}

\newcommand{\elLimitExp}[1][0]{\ensuremath{691\ifthenelse{#1=0}{\GeV}{}}}

\newcommand{\combLimitExp}[1][0]{\ensuremath{739\ifthenelse{#1=0}{\GeV}{}}}

%% file: introduction.tex
\section{Introduction}
\label{section:Introduction}

The large mass of the top quark~\cite{PhysRevD.86.010001} 
may indicate that it is not an elementary particle, 
but has a composite structure, as has been proposed in several 
models of new physics~\cite{Georgi:1994ha,Lillie:2007hd,Pomarol:2008bh,Kumar:2009vs}.
The existence of an excited top quark~($\cPqtstar$) would provide a 
direct test of this possibility~\cite{baur90,harris96}. 
Weak isodoublets can be used to describe both the left-handed and right-handed components 
of a $\cPqtstar$, and provide finite masses prior to the onset of electroweak symmetry 
breaking~\cite{baur90}. 
Thus, in contrast to the heavy top quark of a fourth generation model, 
the existence of an excited top quark is not ruled out by the recent discovery 
of a Higgs boson with properties consistent with those of a 
standard model~(SM) Higgs particle~\cite{Aad:2012tfa,Chatrchyan:2012ufa,Chatrchyan:2013lba}.
It has also been suggested that the top quark may have
higher spin excitations, and in particular,
in string realizations of the 
Randall--Sundrum~(RS) model~\cite{Randall-Sundrum1,Randall-Sundrum2},
the right-handed $\cPqtstar$~quark is expected to be the lightest 
spin-3/2 excited state~\cite{hassanain09}.

This analysis adopts a model in which a $\cPqtstar$~quark has spin~$3/2$
and decays predominantly to a top~quark through the emission of a 
gluon~(\cPg)~\cite{moussallam89,hassanain09,stirling12,Dicus:2012uh}.
A spin-$3/2$ excitation of a spin-$1/2$ quark is governed 
by the Rarita-Schwinger~\cite{Rarita-Schwinger} vector-spinor Lagrangian, 
with the rate of production of spin-$3/2$ quarks being larger than that 
of spin-$1/2$ quarks of similar mass.
This is because the pair production cross section of spin-3/2 quarks is 
proportional to~$\hat{s}^3$ for large values of~$\hat{s}$, while that of 
spin-1/2 quarks is proportional to~$\hat{s}^{-1}$, where~$\hat{s}$ is 
the square of the energy in the parton-parton collision rest frame.
Consequently, at large proton-proton center-of-mass energies~$\sqrt{s}$, integrating 
over parton distribution functions~(PDF), spin-3/2 quarks benefit more 
from contributions at large parton momentum fractions~($x$) than 
spin-1/2 quarks~\cite{moussallam89,hassanain09}.
The growth of the cross section with energy as $\hat{s}^3$ violates unitarity at sufficiently high 
energies, but the relationship is valid at the energies and mass scales accessible at
the CERN Large Hadron Collider~(LHC). 
The $\cPqtstar$ in the RS model is expected to have a pair production 
cross section at $\sqrt{s}=8\TeV$ of the order of a few pb for 
a $\cPqtstar$ of mass $m_{\cPqtstar}=500\GeV$~\cite{stirling12,Dicus:2012uh}.
This cross section is calculated to leading order with a scale $Q=m_{\cPqtstar}$.

Searches have been performed for single production 
of excited generic quarks~($\cPq^*$) that decay to $\cPq\cPg$, a process that dominates in spin-1/2 models.
The Compact Muon Solenoid~(CMS) collaboration has excluded~$\cPq^*$ in the mass range 
of 1\TeV to 3.19\TeV~\cite{Khachatryan:2010jd},
and the ATLAS collaboration has set a lower limit on $m_{\cPq^*}$ of 
$2.83\TeV$~\cite{ATLAS:2012pu}.
However, a $\cPqtstar$~signal would not have been observed in such searches. 
We present the first dedicated search at the LHC for the 
pair production of excited top quarks with spin 3/2 that decay to $\cPqt+\cPg$.

We assume a 100\%~branching fraction for B($\cPqtstar\to\cPqt\cPg$),
the channel that is expected to be the dominant decay mode~\cite{hassanain09,Dicus:2012uh}. 
With mixing between spin-1/2 and spin-3/2 states suppressed, 
the production of mixed pairs of~\cPqt\cPaqtstar{} or~\cPaqt\cPqtstar{}
is expected to have a much smaller cross section than \cPqtstar\cPaqtstar, despite being kinematically favored~\cite{moussallam89,hassanain09}. 
We consider therefore only pair production of the $\cPqtstar$~quark and
its antiparticle, and focus on decay channels containing a single charged lepton~($\ell$) 
specifically in the $\mu+$jets and $\Pe+$jets final states. 
We use a fourth-generation model to mimic the $\cPqtstar$~signal
because the \MADGRAPH~5.1.3.30~\cite{madgraph} Monte Carlo~(MC) generator does not normally include spin-3/2 particles. 
We show in the following section that this choice does not affect the results of the study.

The analysis strategy is to reconstruct the $\cPqtstar$~mass from the 
$\cPqtstar\cPaqtstar\to\cPqt\cPaqt\cPg\cPg\to\PWp\cPqb\PWm\cPaqb\cPg\cPg\to\ell^{+}\nu_{\ell}\cPqb\cPaq\cPq^{\prime}\cPaqb\cPg\cPg$~decay 
chain, including charge-conjugate states, and to compare the resultant mass distributions expected for signal and background.
The analysis is performed using $\Pp\Pp$~collision data at $\sqrt{s}=8\TeV$ collected 
with the CMS detector, corresponding to an integrated luminosity of $\muLumi[1]\pm\muLumiUnc\fbinv$. 

%% file: data.tex
\section{The CMS detector, simulations and data}
\label{sec:data}

The central feature of the CMS apparatus is a
superconducting solenoid of 6\unit{m} internal diameter,
providing a magnetic field of 3.8\unit{T}.
A silicon pixel and strip tracker, a lead-tungstate crystal electromagnetic calorimeter (ECAL),
and a brass/scintillator hadron calorimeter (HCAL) reside within the magnetic volume.
Muons are measured in gas-ionization detectors embedded in the steel flux
return yoke outside of the solenoid.
Extensive forward calorimetry complements the coverage provided by the
central barrel and endcap ECAL and HCAL detectors.
The CMS experiment uses a right-handed coordinate system, with origin at the center of the detector,
the $x$~axis pointing to the center of the LHC ring, the
$y$~axis pointing up~(perpendicular to the plane of the LHC ring), and the $z$~axis
along the counterclockwise beam direction. The polar angle~$\theta$ is measured
from the positive $z$~axis,
and pseudorapidity is defined as ${\eta = -\ln[\tan(\frac{\theta}{2})]}$.
The azimuthal angle~$\phi$ is defined in the $x$-$y$~plane.
A more detailed description of the detector can be found in Ref.~\cite{Chatrchyan:2008zzk}.

The data are collected using single-lepton$+$jets triggers.
The single-muon$+$jets trigger requires that at least one muon candidate is
reconstructed within $\abs{\eta}<2.1$ and has a transverse momentum~$\pt>17\GeV$.
The single-electron$+$jets trigger requires that an electron candidate is
reconstructed with $\pt>25\GeV$ within $\abs{\eta}<2.5$~(with a small region
of exclusion in the transition region between the ECAL barrel and endcaps
at $\abs{\eta}\approx1.5$).
Both channels must have at least three jets reconstructed
within $\abs{\eta}<2.5$ and with transverse momenta larger than a
value which was increased in steps from 20~to 45\GeV, as the average instantaneous luminosity of the LHC increased during the course of data taking.

Simulated inclusive $\cPqtstar\cPaqtstar$ events, including up to two
additional hard partons, are generated for $\cPqtstar$
masses of 450--950\GeV in 50\GeV steps using the
\MADGRAPH~5.1.3.30~\cite{madgraph} event generator and the
CTEQ6L1~PDF~\cite{Pumplin:2002vw}.
We use \PYTHIA~6.426~\cite{pythia} to model parton showers and hadronization.
The generated events are processed through a simulation of the CMS~detector based
on \GEANTfour~4.3.1~\cite{geant4}, and reconstructed using the same algorithms
as used for data.
The \MADGRAPH generator does not normally include spin-3/2 particles,
so we use a fourth-generation model to mimic the $\cPqtstar$~signal.
As our acceptance criteria are not sensitive to opening angles between particles
or other variables that might be affected by spin,
we do not expect this choice to impact our results.
Although it was not possible to simulate all samples this way, to check this
assumption, we were able to include the Rarita--Schwinger Lagrangian
in \MADGRAPH, and generate a true spin-3/2 event sample.
The acceptances for the spin-3/2 and spin-1/2 samples are found to be equal within the uncertainties, which are of order 5\%.
The direction and momentum of jets from final-state particles is
consistent between the two samples, although the number of jets produced
in the spin-3/2 sample is higher than it is in spin-1/2.

Although the analysis is based mainly on an estimate of background obtained from data,
we also use MC simulation of background processes to study the
modeling of the data and to provide a cross-check of our results.
The production of $\ttbar$~events with up to three additional hard partons,
single-top-quark production in the $s$-channel and
$t$-channel, $\cPqt\PW$ processes, $\PW+$jets and $\cPZ+$jets production,
and the smaller diboson~($\PW\PW$, $\PW\cPZ$, $\cPZ\cPZ$), $\ttbar\PW$,
and $\ttbar\cPZ$ contributions have all been modeled in the MC simulation used for these checks.
The diboson processes are generated with the \PYTHIA program,
while the other processes are modeled using the \MADGRAPH package.
The cross section for single top-quark production is taken from Ref.~\cite{Kidonakis:2008mu},
and the cross section for $\PW\cPZ$ production is computed using
the \MCFM~generator~\cite{MCFM1,MCFM2}.
The cross sections for $\ttbar\PW$ and $\ttbar\cPZ$ are computed using \MADGRAPH.
All other cross sections are normalized to the published CMS
measurements~\cite{Chatrchyan:2013oev,Chatrchyan:2013qca}.
All simulated samples include additional contributions from
minimum bias events that model the
energy from overlapping $\Pp\Pp$~collisions
within the same bunch crossing~(``pileup'')
at large instantaneous luminosities.

%% file: selection.tex
\section{Event reconstruction}
\label{sec:reconstruction}

Events are reconstructed using a particle-flow algorithm,
in which each particle is reconstructed and identified by means of an
optimized combination of information from all subdetectors~\cite{CMS-PAS-PFT-09-001}.
The energies of photons are obtained directly from the ECAL signals,
corrected for effects of the algorithm used for noise suppression in the readout.
The energies of electrons are determined from a combination of the track momenta at the
main interaction vertex, the corresponding ECAL cluster energy, and the energy sum of
all bremsstrahlung photons emitted along their trajectories.
The energies of muons are obtained from the corresponding track momenta measured in the silicon tracker and outer muon system.
The energies of charged hadrons are determined similarly from a combination of track momenta
and the corresponding ECAL and HCAL energies, which are corrected for effects of noise suppression.
Finally, the energies of neutral hadrons are obtained from calibrated ECAL
and HCAL energies~\cite{CMS-PAS-PFT-09-001,CMS-PAS-PFT-10-001,CMS-PAS-PFT-10-002,CMS-PAS-PFT-10-003}.

We require events to contain at least one interaction vertex,
with $>10$~associated charged-particle tracks, located within a
longitudinal distance $\abs{z}<24\cm$ and a radial distance $r<2\cm$
from the center of the CMS detector.
The vertex with the largest value for the sum of the $\pt^2$ of
the associated tracks is taken as the primary vertex for the hard collision.

Muon candidates are reconstructed using hits in the silicon tracker
and in the outer muon system
by making a global fit to the hits in both detectors~\cite{Chatrchyan:2012xi}.
Electron candidates are reconstructed from energy clusters
in the ECAL that are also matched to tracks in the tracker.
Trajectories of electron candidates are reconstructed using a CMS
model of electron energy loss, and fitted using a
Gaussian sum filtering algorithm~\cite{electronGSF}.
Jets are reconstructed from particle-flow candidates using the anti-\kt jet
clustering algorithm~\cite{Cacciari:2008gp} with a distance parameter of 0.5,
and jet energies are corrected to establish a uniform relative response
of the calorimeter in $\eta$, and a calibrated absolute response in \pt~\cite{CMS-PAPERS-JME-10-011}.

Jets are identified as originating from a \cPqb~quark through a
combined secondary vertex~(CSV) algorithm~\cite{Chatrchyan:2012jua} that
provides optimal \cPqb-tagging performance.
This algorithm uses a multivariate discriminator to combine information
on the significance of the impact parameter, the jet kinematics,
and the location of the secondary vertex.
The working point of the CSV discriminant is chosen such that light quarks are mistagged at
a rate of 1\%, with a corresponding efficiency for identifying \cPqb-quark jets of 70\%.
Small differences in \cPqb-tagging efficiencies and mistag rates between
data and simulated events are accounted for by scale factors applied to
the simulation.

The imbalance in transverse momentum~(\PTm) of an event is defined as the magnitude of
the vector sum of the transverse momenta of all objects reconstructed using
the particle-flow algorithm.
The corrections applied to jet energies are propagated to the measured~\PTm.

\section{Offline event selection}
\label{sec:selection}

Charged leptons from $\cPqt\to\cPqb\ell\nu$ decays are expected to be isolated from nearby jets.
Relative isolation,~$I$, is defined as the ratio of the scalar sum of the transverse momenta
of all photons, charged hadrons, and neutral hadrons, associated with the primary vertex,
in an angular cone around the lepton direction to the lepton~\pt.
The sum includes all these particle-flow candidates within a cone of
$\Delta R\equiv\sqrt{(\Delta\eta)^2+(\Delta\phi)^2}<0.4$ around the muon candidate, and~${<}0.3$
around the electron candidate, where $\Delta\eta$ and $\Delta\phi$ are the differences
in pseudorapidity and azimuth relative to the lepton direction.
Estimates of the contributions from pileup
interactions to the neutral hadron
and photon energy components are subtracted from the
above sums~\cite{Chatrchyan:2012xi,Cacciari:2007fd}.

Event candidates in the $\mu+$jets channel are required to have only one
muon with $\pt>26\GeV$, $\abs{\eta}<2.1$, $I<0.12$,
and with transverse and longitudinal distances of closest approach
to the primary vertex of $d_r<2\mm$ and $|d_z|<5\mm$, respectively.
Candidates in the $\Pe+$jets channel are required to have only one electron
with $\pt>30\GeV$, $\abs{\eta}<1.44$~(restricting electrons to the central rather than forward
regions reduces contributions from generic multijet events),
$I<0.1$, and $d_r<0.2\mm$.
These selections are more restrictive than those used for the trigger,
ensuring the selected leptons are in the plateau of the trigger efficiency.

Additional selection criteria require at least six jets with $\pt>30\GeV$
and $\abs{\eta}<2.5$.
To ensure high trigger efficiency, the three leading jets~(i.e. with largest \pt) are
each required to have $\pt>45\GeV$ in the initial data-taking period,
and $\pt>55$,~$45$, and~$35\GeV$, respectively, in the subsequent data-taking periods.
At least one jet must be \cPqb-tagged through the CSV algorithm.
In the region of acceptance, the loss of efficiency arising from the turn-on of the 
acceptance as a function of jet-\pt is very small~(less than 1\%), and the total trigger 
efficiency ranges between 85~and 100\%.
For the signal, the average efficiency is ${\approx}91\%$, while for the background it is ${\approx}90\%$.

\begin{figure}[hbtp]
  \centering
  \begin{tabular}{cc}
      \includegraphics[width=0.45\textwidth]{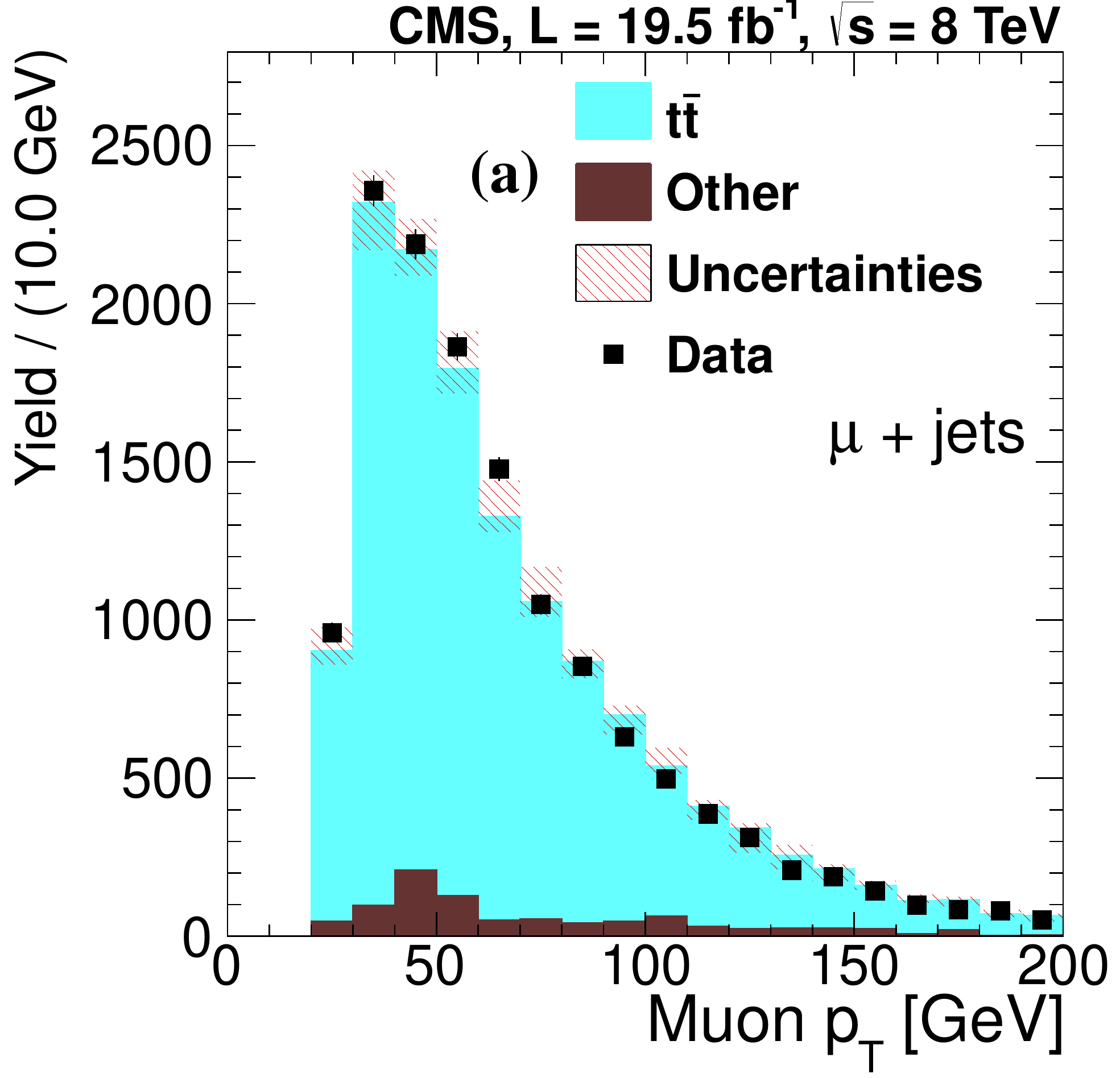}    &
      \includegraphics[width=0.45\textwidth]{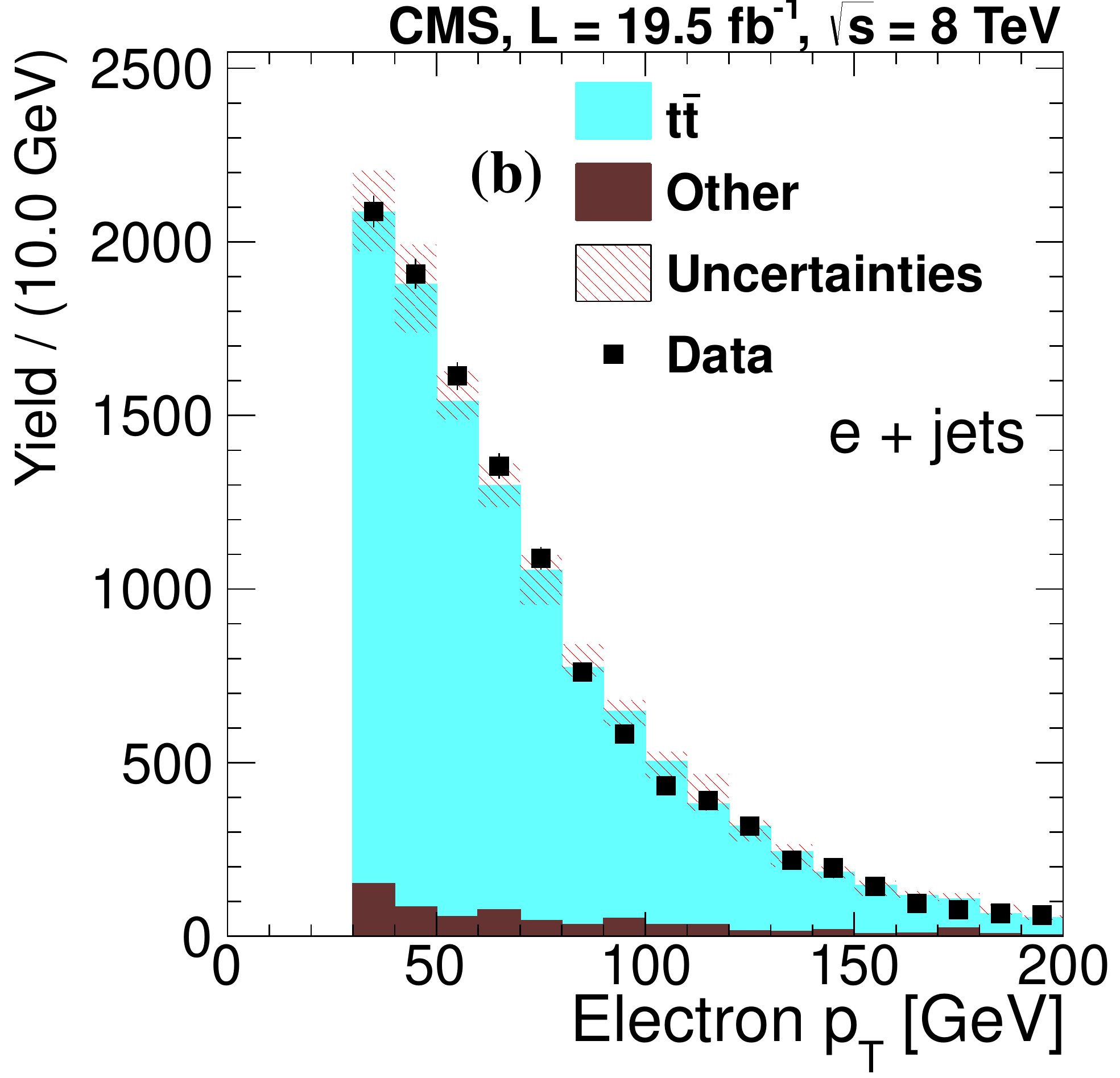}    \\
      \includegraphics[width=0.45\textwidth]{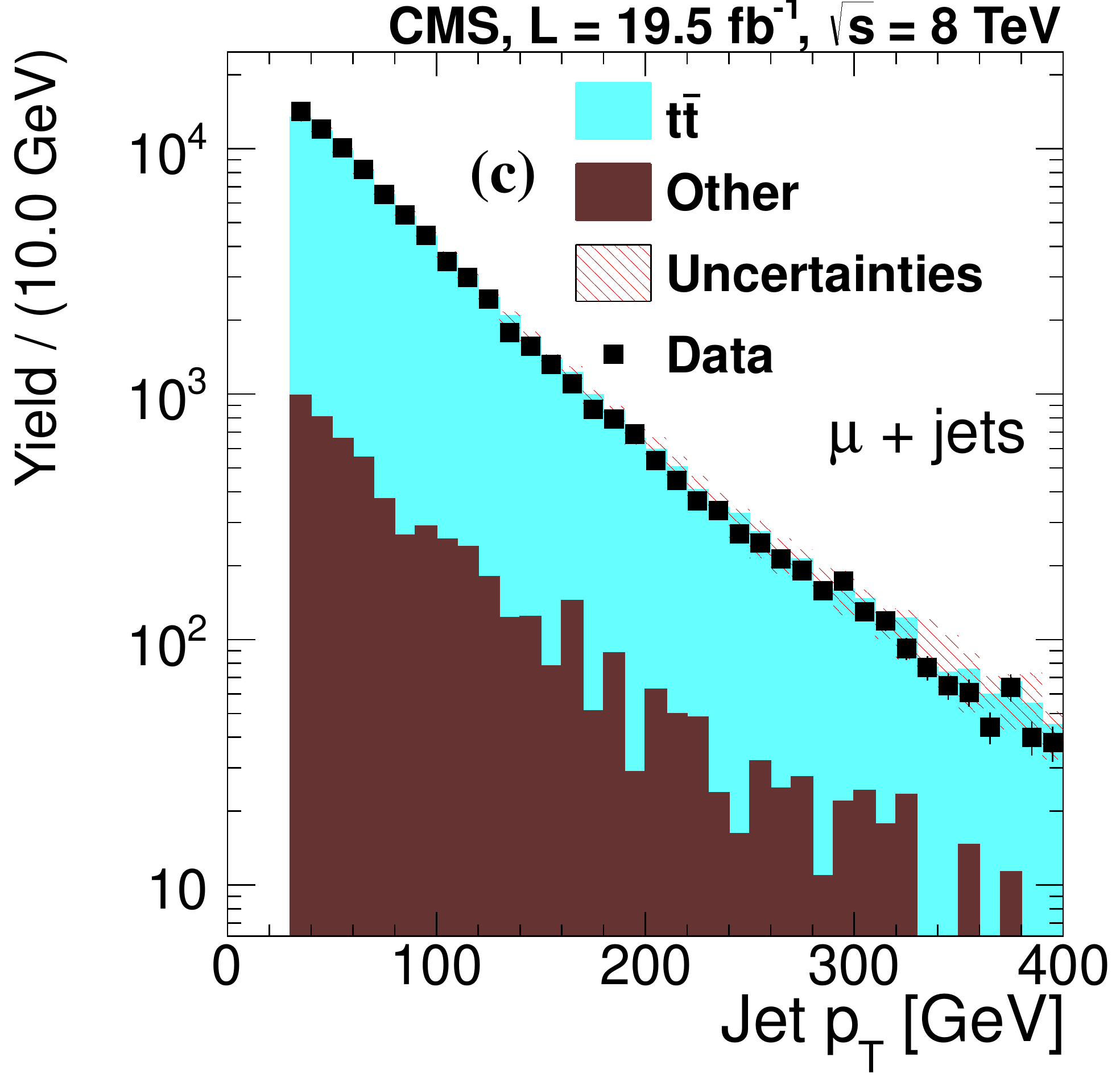} &
      \includegraphics[width=0.45\textwidth]{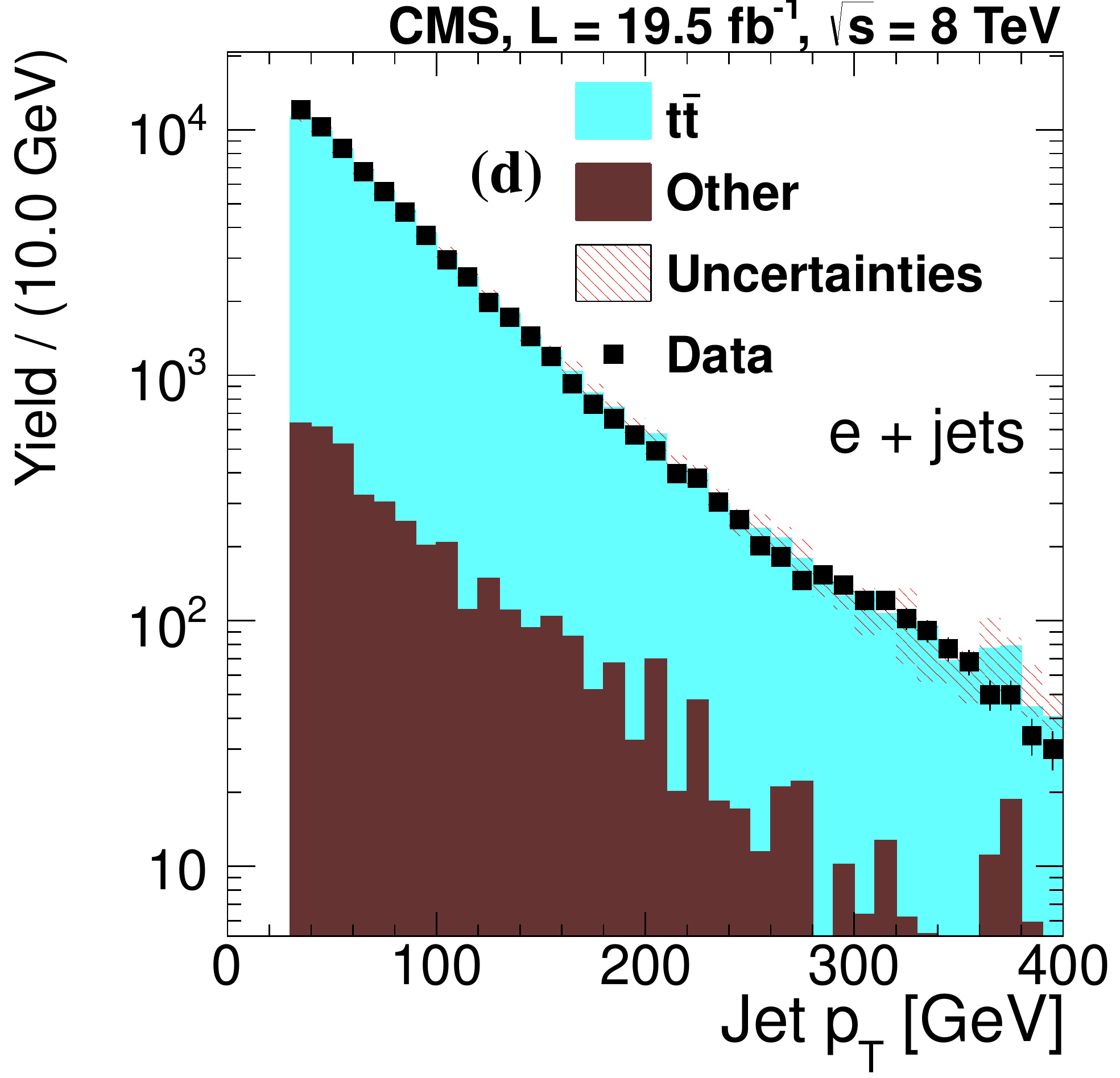} \\
      \includegraphics[width=0.45\textwidth]{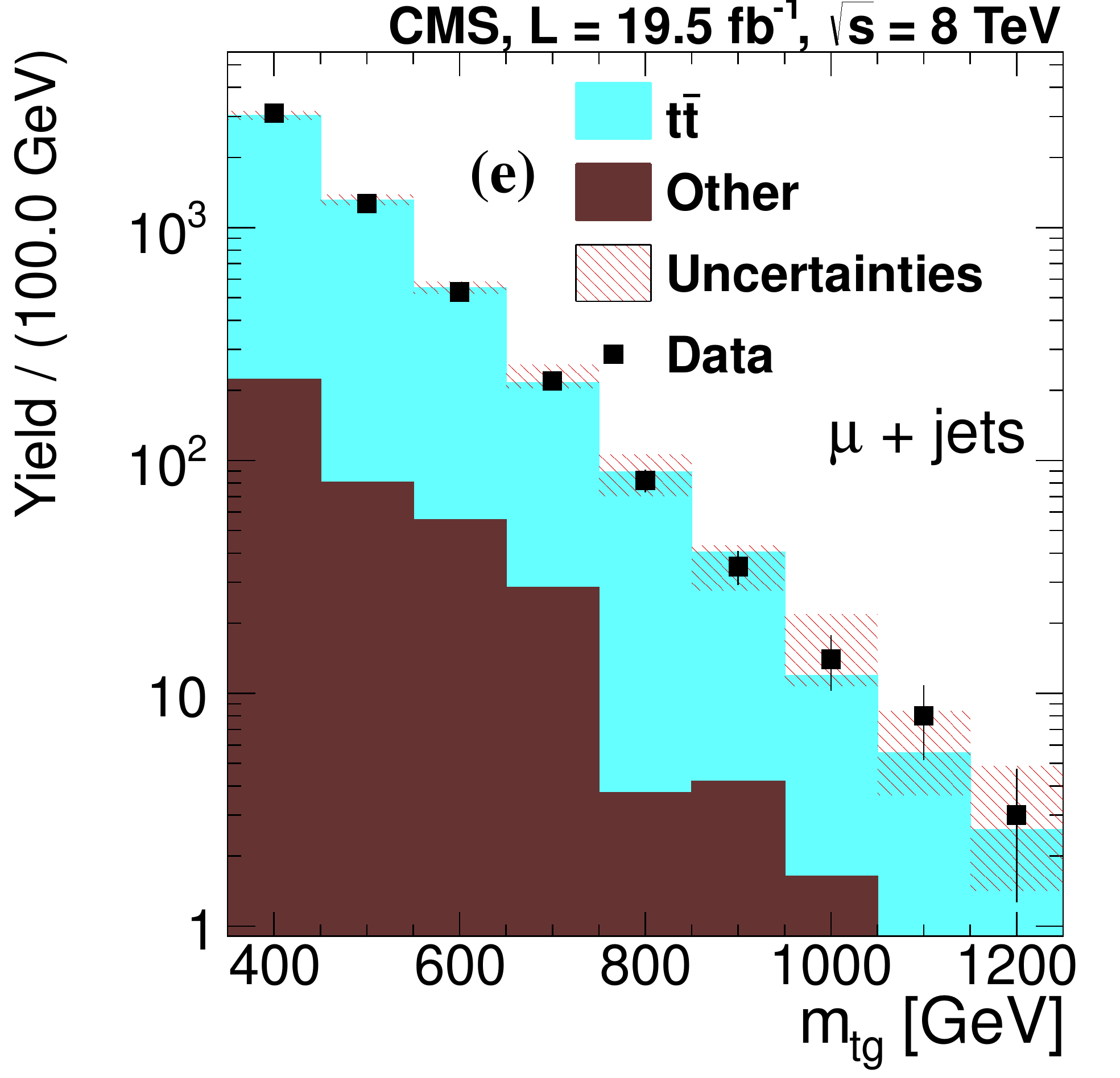} &
      \includegraphics[width=0.45\textwidth]{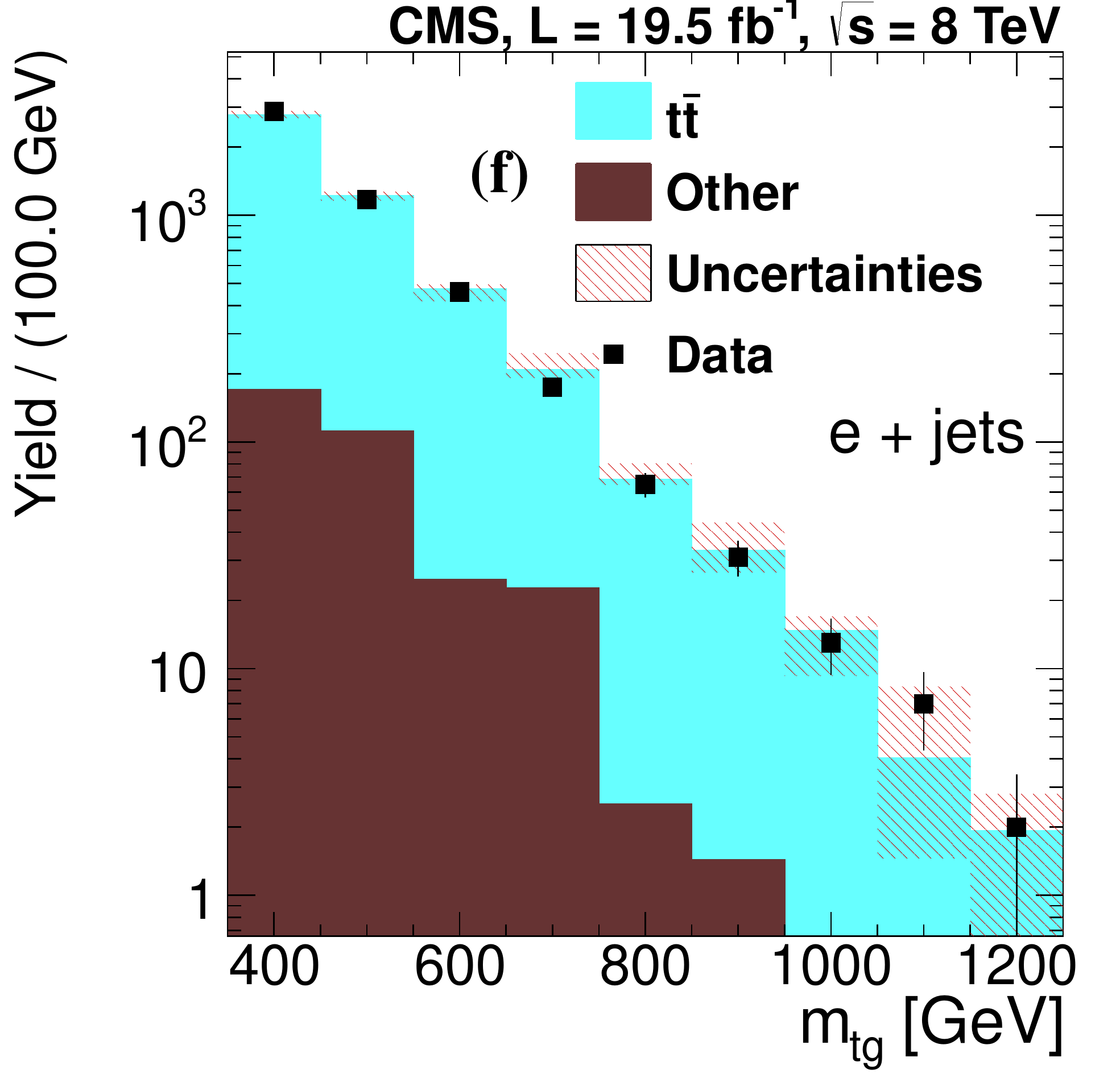}
  \end{tabular}
  \caption{Kinematic distributions of single $\ell\,+\!>5$-jet events
    in data~(points), compared to MC simulation
    normalized to the number of events observed in data.
    Shown are \pt spectra for muons~(a) and electrons~(b),
    and jet spectra for the channels $\mu+$jets~(c) and $\Pe+$jets~(d).
    The reconstructed $m_{\cPqt\cPg}$ distribution is shown for the
    $\mu+$jets channel in~(e) and for $\Pe+$jets in~(f).
  }
  \label{fig:distributions}
\end{figure}

Signal events pass our selections with efficiencies varying
from 18\% at low $\cPqtstar$~masses to 20\% at higher masses.
The largest efficiency losses arise from the lepton isolation
and jet requirements.
After the application of all selection criteria, we observe \muYield~events in
the $\mu+$jets channel and \elYield~events in the $\Pe+$jets channel.
The yields predicted from simulated SM background processes are
\muYieldPred~events in the $\mu+$jets channel and
\elYieldPred~events in the $\Pe+$jets channel.
The event yield uncertainties are dominated by uncertainties in the choice of the renormalization and factorization
scales used in the \MADGRAPH generation of \ttbar~events, and by the uncertainty in the jet energy scale~(JES).
The small deficits in data relative to SM expectations are
within the estimated uncertainties.
Furthermore, the differential distributions of the kinematic variables are in agreement.
We determine this by renormalizing the simulation to the number of events observed in data,
and find agreement in the distributions of all kinematic
variables for the predicted and observed $\ttbar$~events, as seen in Figure~\ref{fig:distributions}.
Of particular importance, the distribution in the mass of the
$\cPqt\cPg$~system~(see Section~\ref{sec:mass} for details) is reproduced by the simulation.
In the following sections, we describe the strategy adopted for reconstructing the mass of
the $\cPqtstar$~candidate and for estimating the background from control samples in data.

%% file: reconstruction.tex
\section{Mass reconstruction}
\label{sec:mass}

The dominant background to a $\cPqtstar\cPaqtstar$~signal is
expected to be from SM $\ttbar$~production in association with extra jets. %
We therefore use the reconstructed mass distribution of the
$\cPqt+$jet systems to distinguish a $\cPqtstar\cPaqtstar$~signal from
$\ttbar$~background.

The procedure adopted for reconstructing the mass is as follows.
In the $\ell+$jets channels, one \PW~boson decays leptonically,
while the other decays into a $\cPq^{\prime}\cPaq$~pair, i.e.
$\cPqtstar\cPaqtstar\to(\ell\nu\cPqb\cPg)(\cPq^{\prime}\cPaq\cPqb\cPg)$.
The reconstructed objects in the event, namely, the charged lepton, the \PTm,
and the six leading jets correspond to the particles in the decay of the $\cPqtstar\cPaqtstar$~system,
and are assigned to one of the initially produced objects.
We assume that the \PTm is carried away entirely by the neutrino emitted by
the leptonically decaying $\PW$~boson.
The longitudinal component of the neutrino momentum~($p_{\mathrm{z}}$) cannot be measured,
but an initial estimate of its value is determined~(within a two-fold ambiguity)
using the requirement that the two reconstructed top~quarks have the same mass.
All possible permutations of jet-parton assignments are considered in the analysis,
subject to the condition that a \cPqb-tagged jet must be assigned to one of the
\cPqb~quarks.
When multiple jets are \cPqb-tagged, all binary combinations are interpreted as \cPqb~quarks.

After assigning the reconstructed objects to their progenitor particles,
a constrained kinematic fit is performed to the $\cPqtstar\cPaqtstar$~hypothesis
to improve the resolution of the reconstructed mass of the $\cPqtstar$~candidates.
We use an algorithm originally designed to measure
$m_{\cPqt}$ in $\ttbar$~events~\cite{Chatrchyan:2012cz,Abbott:1998dc},
but modified to reconstruct $\cPqtstar\cPaqtstar$~events that contain two additional jets.
The momenta of the reconstructed objects are adjusted in the fit to simultaneously
satisfy the following constraints:
\begin{alignat}{3}
m(\ell\nu) &= m(\cPq\overline{\cPq})& &= m_{\PW}, \label{eq:mw}\\
m(\ell\nu\cPqb) &= m(\cPq\overline{\cPq}{\cPqb})& &= m_{\cPqt}, \label{eq:mt}\\
m(\ell\nu\cPqb\cPg) &= m(\cPq\overline{\cPq}\cPqb\cPg)& &= m_{\cPqt\cPg}, \label{eq:mtstar}
\end{alignat}
where $m_{\PW}=80.4\GeV$ is the mass of the \PW~boson,
$m_{\cPqt}=173.5\GeV$ is the mass of the top quark~\cite{PhysRevD.86.010001},
and $m_{\cPqt\cPg}$ is a free parameter,
the resolution of which is improved through the fit.

All the momentum components of the reconstructed objects,
with the exception of $p_{\mathrm{z}}$ of the neutrino momentum, are measured.
There is consequently one unknown and seven constraints to the kinematics:
(i)~two from each of Equations~(\ref{eq:mw}) and~(\ref{eq:mt}),
(ii) two from the conservation of transverse momentum in the collision,
and (iii) one constraint from Equation~(\ref{eq:mtstar}).
We perform a fit to the $\cPqtstar\cPaqtstar$~hypothesis by minimizing
a $\chi^2$ computed from the sum of the squares of the difference
between the measured components of momenta of all reconstructed objects
and their fitted values, each term divided by the sum of the squares of their estimated
uncertainties, subject to the remaining six constraints.
The jet permutation with the smallest $\chi^2$~value
is chosen to represent the event.

The above procedure selects the correct jet-parton assignment in
about 11\% of the simulated $\cPqtstar\cPaqtstar$~events,
with the $\cPqtstar$~quark that decays through the $\PW\to\ell\nu_{\ell}$~mode
being reconstructed correctly in about 1/3 of the lepton+jets final states.
We have studied the possibility of including up to eight jets in the
reconstruction~(\ie considering all combinations of six out of the
leading six, seven, or eight jets).
However, there is little gain using this approach,
despite that it yields 13\% in correct assignments.
A major reason for getting the wrong jet-parton combination is that
in approximately 40\% of the $\cPqtstar\cPaqtstar$ events, at least
one jet from the $\PW\to\cPq^{\prime}\cPaq$~decay
fails the offline jet-\pt requirement.
In events where all the hadronic decay products are included among the six
leading jets, the correct jet-parton assignment is selected 68\% of
the time, but this fraction decreases significantly if we consider
up to eight jets in the final state.
Consequently, $\chi^2$ fits using more than six jets contain far more background.
Variations in the fraction of events with correct jet-parton assignments do not significantly affect our final results~(discussed in Section~\ref{sec:analysis}).
A comparison of the reconstructed $\cPqtstar$~mass distributions obtained for the spin-1/2 
and spin-3/2 samples, using the kinematic fit, reveals no significant dependence on the spin.

%% file: background.tex
\section{Background model and extraction of \texorpdfstring{$\cPqtstar$}{t*} signal}
\label{sec:background}

We model the $m_{\cPqt\cPg}$ distribution for the background
from the SM using a Fermi function:
\begin{equation}
f(m) = \frac{a}{1+\re^{\frac{m-b}{c}}},
\label{fn:1stfunction}
\end{equation}
where $m$ represents the mass reconstructed under the
$\cPqtstar$~hypothesis, and $a$, $b$, and $c$ are parameters
that are determined through a fit to the data.
The $m_{\cPqt\cPg}$ distribution for a $\cPqtstar\cPaqtstar$~signal
is taken from simulated events.

\begin{figure}[hbtp]
\centering
\begin{tabular}{cc}
  \includegraphics[width=0.48\textwidth]{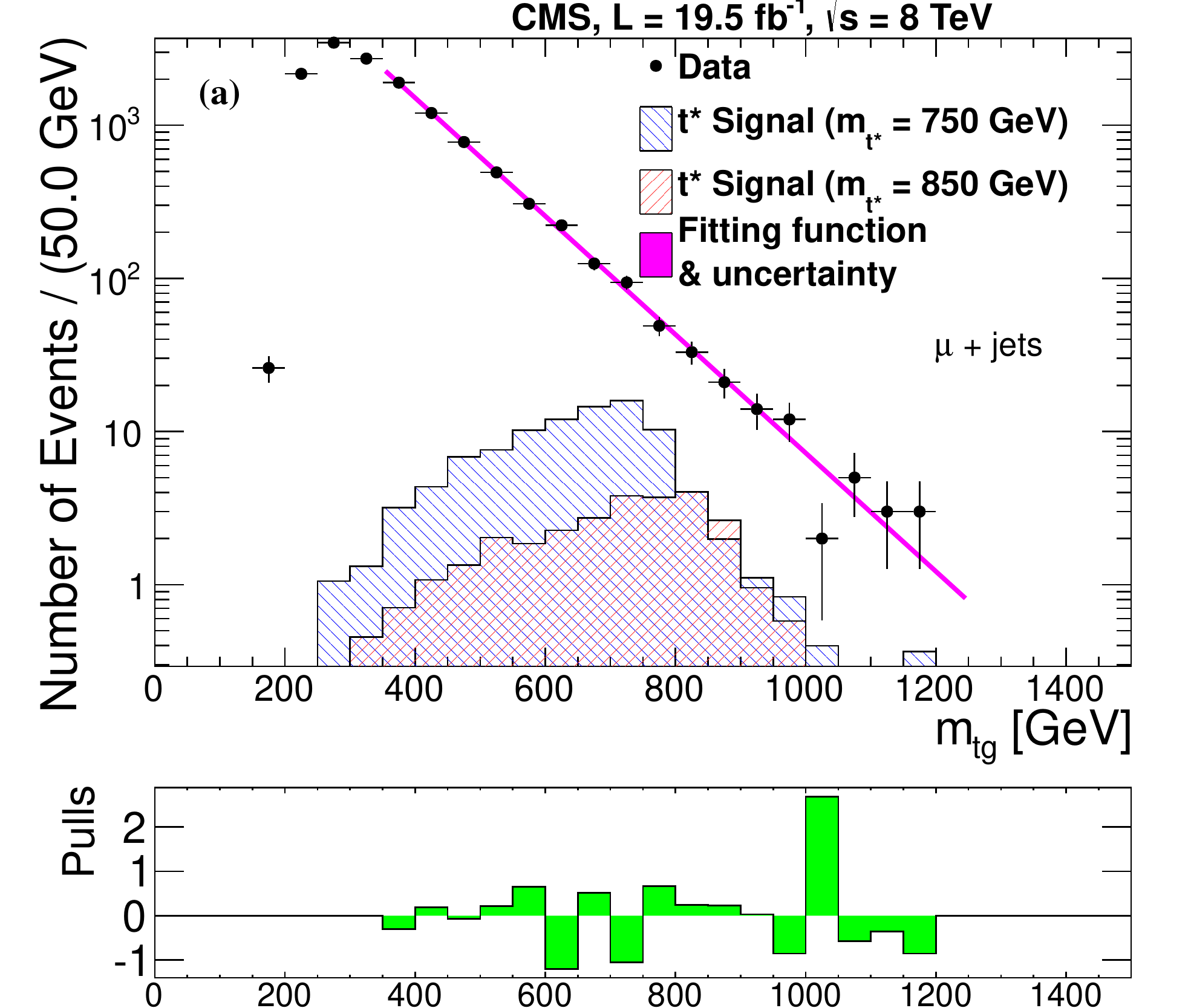} &
  \includegraphics[width=0.48\textwidth]{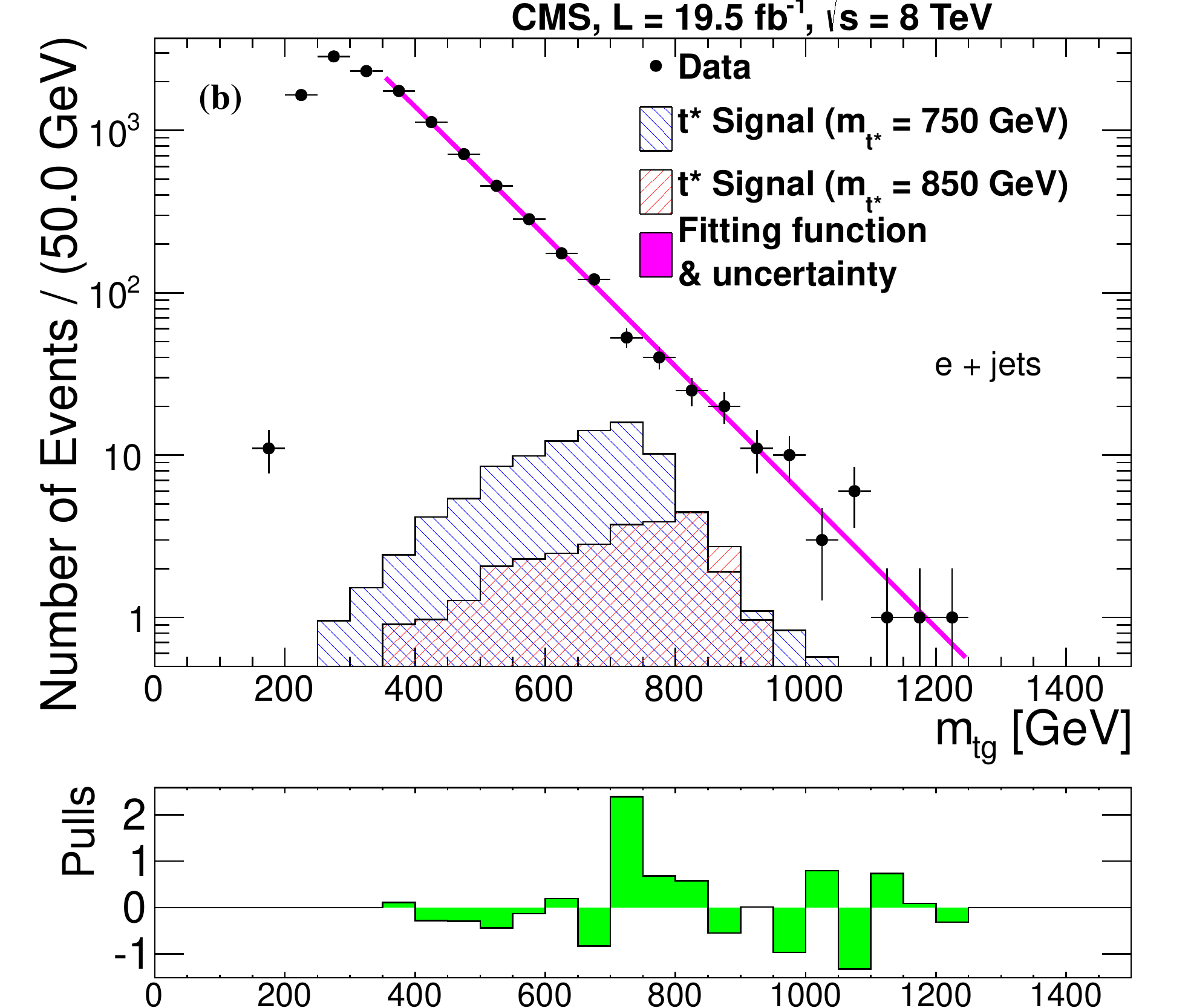} \\
\end{tabular}
\caption{Reconstructed mass spectrum for the $\cPqt\cPg$~system in data~(points),
along with a fit of the background~$f(m)$ of Equation~(\ref{fn:1stfunction}) to the data
in the $\mu+$jets channel~(a) and $\Pe+$jets channel~(b).
The reconstructed masses correspond to the results of kinematic fits for
the jet-quark assignments
that provide the best match to the $\cPqtstar\cPaqtstar$~hypothesis.
Also shown are the expectations of $\cPqtstar$~signals
for $m_{\cPqtstar}=750$~and~$850\GeV$ normalized to the integrated luminosity
of the data.
The lower panels show the ``pulls'' (the differences between observation and expectation, divided by the uncertainty in the expected values).
}
\label{fig:topgluon_mass}
\end{figure}

The $\cPqtstar\cPaqtstar$~signal and the background contributions
in data are estimated simultaneously.
For each generated $m_{\cPqtstar}$~value, we perform a
binned likelihood fit to the sum of the background
function~$f(m)$ and the reconstructed mass spectrum for the
$\cPqtstar\cPaqtstar$~model for $m_{\cPqt\cPg}>350\GeV$.
The $\cPqtstar\cPaqtstar$~cross section and the three parameters
of the background function are varied in this fit.
Figure~\ref{fig:topgluon_mass} shows the distribution of the
reconstructed $m_{\cPqt\cPg}$ for the $\mu+$jets channel~(a)
and $\Pe+$jets channel~(b),
along with the fit to the background.
The small differences between observation and expectation, divided by the uncertainty in the expected values, 
shown below the $m_{\cPqt\cPg}$~distributions, demonstrate that the fitting function describes the background well.
The function~$f(m)$ shown in the figure
represents the contribution from background events only,
and does not include the $m_{\cPqtstar}=750\GeV$ or
$m_{\cPqtstar}=850\GeV$~signals, which are shown separately.

As a check of the stability of background fit, we also performed the fit for $m_{\cPqt\cPg}>650\GeV$ and for $m_{\cPqt\cPg}>850\GeV$.
For each fit we calculated the integral and uncertainty in the function for the range $850\GeV<m_{\cPqt\cPg}<1250\GeV$.
For the nominal range of~$m_{\cPqt\cPg}>350\GeV$, the results are $60.1\pm3.5$ and $46.6\pm3.1$ for the $\mu+$jets and $\Pe+$jets channels, respectively. 
For the ranges $m_{\cPqt\cPg}>650\GeV$ and $m_{\cPqt\cPg}>850\GeV$, the results are in close agreement, 
with values of $59.8\pm5.7$ and $51.2\pm5.5$ for the $\mu+$jets and $\Pe+$jets channels, respectively, for the range $m_{\cPqt\cPg}>650\GeV$ and
$60.3\pm7.8$ and $53.3\pm7.5$ for the range $m_{\cPqt\cPg}>850\GeV$.

To show that the fitting method is sensitive to the presence of
$\cPqtstar$~signal, pseudo-data are generated according
to a probability distribution function representing the sum of
$f(m)$ and a specific $\cPqtstar$~signal.
Performing the kinematic fit on the pseudo-data
provides a cross section for the extracted $\cPqtstar$~signal that indicates
no bias in the fitting procedure.

As a check of our method, we also model the background using MC samples.
As noted in Section~\ref{sec:selection}, the distribution of the
simulated background samples is in agreement with the data.
The background and signal MC templates are fit to the data to determine
their contributions.

%% file: uncertainties.tex
\section{Systematic uncertainties}
\label{sec:uncertainties}

Systematic uncertainties influence the assessment of whether the
$m_{\cPqt\cPg}$ distributions for the observed events are
consistent with the presence of a signal,
or with expectations from background alone.
The dominant sources of systematic uncertainty are described below.

The uncertainties in the differential distributions for background are
estimated from the uncertainties in the fitted parameters
of Equation~(\ref{fn:1stfunction}), and incorporated into the calculation of limits,
as discussed in Section~\ref{sec:analysis}.
These uncertainties affect both the distribution and the normalization
of the background.
To determine the overall effect of these uncertainties, we perform
limit setting calculations including and excluding the uncertainties
and find a 5\% effect on the mass limit from the uncertainty in background.

Given that the distributions of signal are based on simulation,
we consider the impact of both experimental and theoretical
sources of uncertainty.
For each source, we adjust the relevant parameters in the simulation
to produce alternative templates for signal.
We take the relative differences between the templates for the
alternative parameters and the templates produced using their
nominal values to estimate the magnitude of the uncertainties
in the final result.
We also consider the effect of uncertainties in the differential
distribution of the signal.
These effects are small, as the mass reconstruction algorithm
tends to change the particle momenta to meet
the kinematic constraints and, in so doing, maintains the stability
of the differential spectra.

The signal is affected by a variety of experimental sources of uncertainty.
The integrated luminosity is known to a precision of
2.6\%~\cite{CMS-PAS-LUM-13-001}.
All jet energies are corrected using standard CMS JES
constants~\cite{CMS-PAPERS-JME-10-011}.
We generate alternative distributions  in $m_{\cPqt\cPg}$ after
rescaling the nominal jet energies by ${\pm}1$~standard deviation,
using the known parametrization of these uncertainties as a function
of jet~\pt and~$\eta$~\cite{CMS-PAPERS-JME-10-011}.
This rescaling is also propagated to the \PTm.
An observed difference in the jet energy resolution~(JER) in
simulation relative to data is taken into account by applying
an $\eta$-dependent \pt smearing of 5--12\% to
the simulated jets, as required to match the measured resolution.
The uncertainty affecting this extra correction is propagated to
the expected $m_{\cPqt\cPg}$ in a way similar to that used for the
jet energy scale.
The uncertainties from \PTm are mostly included in the uncertainties in
the jet energies. We also consider the uncertainty in any remaining
``unclustered energy'' not arising from one of the jets or lepton in the event,
and find that its impact is negligible.
Other sources of experimental uncertainty include those in trigger
efficiencies and corrections to lepton identification efficiencies,
which are measured using
``tag-and-probe'' methods~\cite{Khachatryan:2010xn} in
the data and in simulation.
The systematic uncertainty in \cPqb-tagging efficiency is estimated
by changing the tagging and misidentification rates for
\cPqb, \cPqc, or light-flavor jets according to the uncertainties
estimated from data~\cite{Chatrchyan:2012jua}.
The systematic uncertainty from the modeling of pileup events is
checked by changing the minimum-bias cross section by
${\pm}1$~standard deviation, which changes the
average number of pileup events by ${\pm}4$\%\cite{CMS-PAS-LUM-13-001}.

We estimate the effect of theoretical uncertainties arising from the
choice of PDF by changing the CTEQ PDF parameters within their
estimated uncertainties,  and measuring the effect on the
simulated acceptance.
We further check that a change of the renormalization and
factorization scales from their nominal values
has negligible impact on the signal.

The statistical uncertainties associated with the simulated samples are
also taken into account as a systematic uncertainty in the measurement.
Table~\ref{tab:signal_uncertainties} quantifies the uncertainties
in the normalization of the signal from each of the above sources.
As can be seen from the table, the luminosity and JES uncertainties generally
dominate the overall signal uncertainty.
Nevertheless, the uncertainties in the signal have less than 1\% effect on the limit
while those in the differential distribution of
$m_{\cPqt\cPg}$ for the background have a 5\% impact on the limit.

\begin{table}[htbp]
\centering
\topcaption{Systematic uncertainties in the normalization of the
  $\cPqtstar\cPaqtstar$ templates.
  The specified ranges indicate the minimum and maximum
  uncertainties for the examined values of $m_{\cPqtstar}$.
}
\label{tab:signal_uncertainties}
\begin{tabular}{lcc}
\hline
Source             & $\mu+$jets & $\Pe+$jets \\
\hline
Luminosity         &  2.6\%       &  2.6\%        \\
JES                &  2.3--3.9\%  &  2.2--4.1\%   \\
JER                &  ${<}1$\%      &  ${<}1$\%       \\
Trigger efficiency &  1.0\%       &  1.0\%        \\
Lepton efficiency  &  0.9--1.3\%  &  $<1$\%       \\
\cPqb-tagging      &  0.6--1.5\%  &  0.8--1.4\%   \\
Pileup             &  ${<}1$\%      &  ${<}1$\%       \\
PDF                &  0.3--1.9\%  &  1.3--1.9\%   \\
MC statistics      &  1.9\%       &  2.0\%        \\
\hline
\end{tabular}
\end{table}

%% file: analysis.tex
\section {Statistical analysis and extraction of limits}
\label{sec:analysis}

We examine the top$+$jet mass spectrum for evidence of
$\cPqtstar$~quark decay into the top$+$gluon final state.
The $\cPqtstar\cPaqtstar$~cross section determined by the fit
described in Section~\ref{sec:background} is consistent with no
signal for each tested value of $m_{\cPqtstar}$.
In the absence of evidence for any excess, we set an upper bound
on the inclusive $\cPqtstar\cPaqtstar$~production
cross section~($\sigma$) using Bayesian
statistics~\cite{PhysRevD.86.010001}, and a uniform prior for
a cross section of $\sigma>0$.
The systematic uncertainties for signal are included through
``nuisance'' parameters assuming log-normal priors that are integrated
over in the process of computing the likelihood~\cite{Prosper:1988ah}.
The combination of the function~$f(m)$ for background and a
template for signal is used in a log-likelihood
fit to the data.
The uncertainty in the differential distribution for the background
is incorporated by integrating over the parameters of the fitted
background assuming uniform priors.
The integration over such nuisance parameters is performed over
a sufficiently large range around the best-fit values to ensure
that the results are stable.
To combine the $\mu+$jets and $\Pe+$jets channels,
we multiply the likelihoods for the two sets of lepton events.
Many of the uncertainties are correlated between the two channels,
and accounted for by requiring the corresponding nuisance
parameters to have the same value in both channels.
Expected limits are obtained by generating pseudo-experiments
based on the fitted $f(m)$~(ignoring $\cPqtstar$~signal),
including the uncertainties on the fit,
and repeating the above calculations as a function of $m_{\cPqtstar}$.

\begin{figure}[hbtp]
\centering
\includegraphics[width=0.7\textwidth]{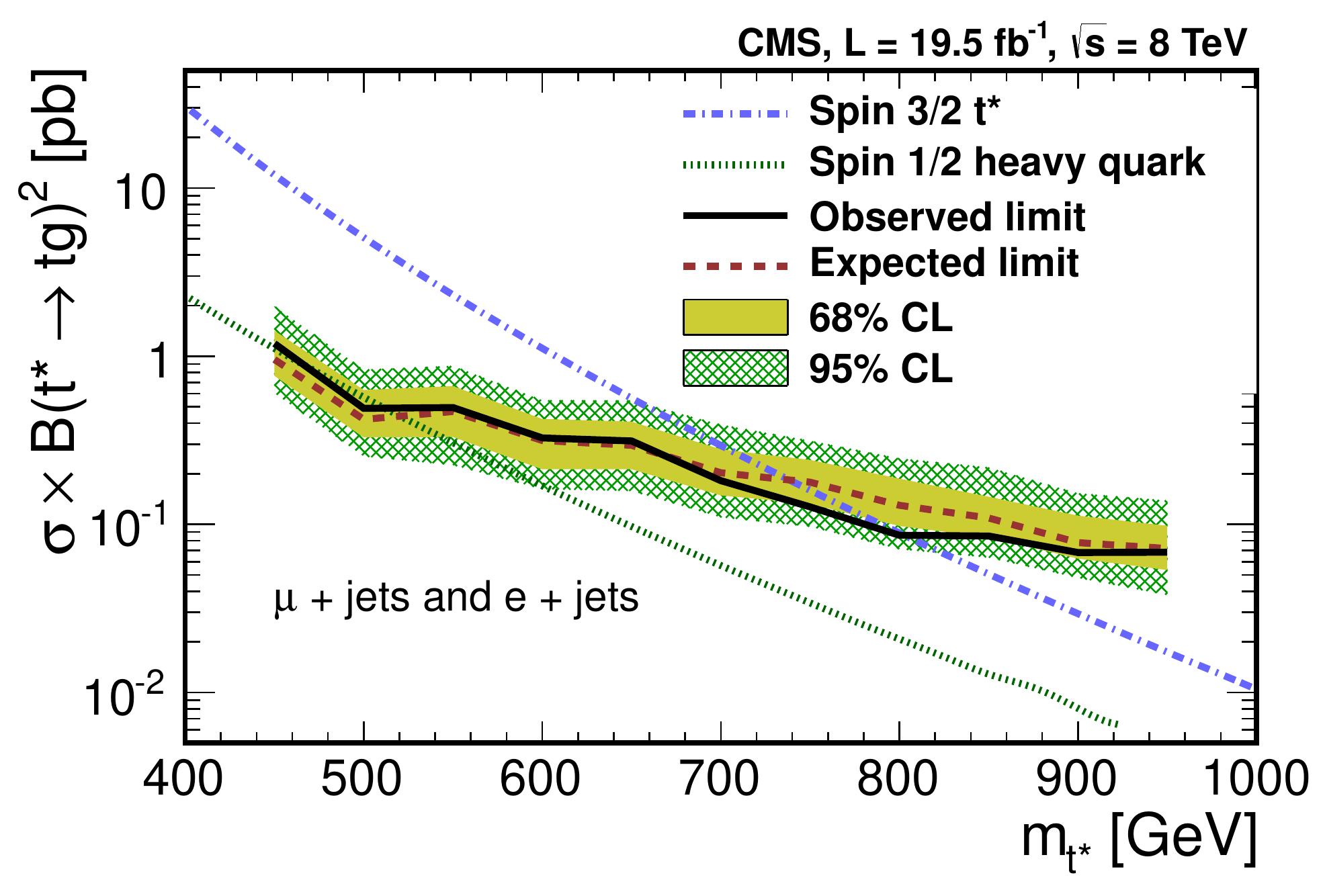}
\caption {The observed (solid line) and expected (dashed line)
  95\%~CL upper limits for the product of the inclusive $\cPqtstar\cPaqtstar$~production cross section
  and the branching fraction $B(\cPqtstar\to\cPqt\cPg)$,
  as a function of the $\cPqtstar$~mass, for the combined lepton data.
  The ranges for ${\pm}1$ and ${\pm}2$ standard deviations for the
  expected limits are shown by the bands.
  The theoretical cross section for the spin-3/2 model is shown by
  the dashed-dotted line~\cite{stirling12}.
  Also shown is the theoretical cross section for producing
  an excited top-quark pair of spin-1/2~\cite{Aliev:2010zk}.}
\label{fig:limit_plot}
\end{figure}

Figure~\ref{fig:limit_plot} shows the observed and expected upper
limits at 95\%~confidence level~(CL) for
the $\cPqtstar$-pair production cross section multiplied by its branching fraction into $\cPqt+\cPg$,
as a function of $m_{\cPqtstar}$.
The lower limit for $m_{\cPqtstar}$ is given by the value at
which the upper limit intersects the leading-order spin-3/2 cross section
from Ref.~\cite{stirling12}.
This procedure yields an observed lower limit for
$m_{\cPqtstar}$ of \combLimitObs{} for the combined muon
and electron data, at 95\%~CL.
The expected limit from pseudo-experiments is $\combLimitExp[1]\GeV$.
The limits are also listed separately for each channel in
Table~\ref{tab:limits}.
\begin{table}[htbp]
\centering
\topcaption{Expected and observed lower limits on $m_{\cPqtstar}$~(\GeVns{}) for a spin-3/2 $\cPqtstar$.}
\label{tab:limits}
\begin{tabular}{lcc}
\hline
Channel    & Expected                 & Observed         \\ \hline
$\mu+$jets & ${\muLimitExp[1]}$       & \muLimitObs[1]   \\
$\Pe+$jets & ${\elLimitExp[1]}$       & \elLimitObs[1]   \\ \hline
Combined   & ${\combLimitExp[1]}$     & \combLimitObs[1] \\
\hline
\end{tabular}
\end{table}
It should be noted that in extracting the lower limit on $m_{\cPqtstar}$, the uncertainties associated with the calculation of the theoretical curve have not been included.
Neglect of the $K$-factor expected from extending the calculation to next-to-leading order implies that the quoted limit is conservative ($K=1.8$ for $\cPqt\cPaqt\cPqb\cPaqb$~production at 14\TeV~\cite{Bevilacqua:2009zn}),
although changing the choice of QCD scale from the assumed value of $m_{\cPqtstar}$ to $2m_{\cPqtstar}$ would decrease the cross section by a factor of $\approx$1.7.

Although not the primary issue under consideration,
Figure~\ref{fig:limit_plot} also shows the limits set for a
spin-1/2 excited quark, based on the next-to-next-to-leading-order cross section calculated with the \textsc{hathor}~(1.5) program~\cite{Aliev:2010zk}.
Assuming the same signature for the decays of excited spin-1/2
and spin-3/2 top quarks,
the expected lower limit on $m_{\cPqtstar}$ for a
spin-1/2 excited quark is 521\GeV, at 95\%~CL.
We exclude such quarks for masses $465<m_{\cPqtstar}<512\GeV$ at 95\%~confidence.

The stability of the limit against changes in the shape of the $m_{\cPqt\cPg}$ distribution, 
due to signal events that are reconstructed using jets not from the decay of a $\cPqtstar$, 
is tested by breaking the signal template into components depending on the number of leading jets that come from a $\cPqtstar$~decay. 
The components are varied by an amount appropriate from initial-state radiation variations and the limit recalculated. 
The limit is found to be stable under these variations.

As noted in Section~\ref{sec:background}, we check the data-driven
method by repeating the analysis using simulated distributions to
represent the background.
The limits obtained using this background estimation agree
with our main result within the assigned uncertainties.

%% file: summary.tex
\section{Summary}
\label{sec:summary}

We have conducted a search for excited spin-3/2 top quarks~($\cPqtstar$) 
that are pair produced in \Pp\Pp~interactions, with each $\cPqtstar$ 
decaying exclusively to a standard model top~quark and a gluon.
Events that have a single muon or electron, and at least six jets, at least 
one of which is identified as a \cPqb-jet, are selected for analysis.
Assuming $\cPqtstar\cPaqtstar$~production, a kinematic fit is 
performed to final-state objects to reconstruct 
$\cPqtstar$~candidates in each event.
The observed mass spectrum of the $\cPqt$-jet system, showing 
no significant deviation from predictions of the standard model, is used 
to set upper limits on the production of $\cPqtstar\cPaqtstar$ 
as a function of the $\cPqtstar$~mass.
By comparing the results with expectations for spin-$3/2$ 
excited top quarks in an extension of the Randall--Sundrum model~\cite{hassanain09},
we exclude $\cPqtstar$~masses below \combLimitObs{} at 95\%~confidence.
This is the first dedicated search for an excited spin-3/2 top quark, 
and sets strong bounds on its existence.

\section*{Acknowledgments}

We congratulate our colleagues in the CERN accelerator departments for the excellent performance of the LHC and thank the technical and administrative staffs at CERN and at other CMS institutes for their contributions to the success of the CMS effort. In addition, we gratefully acknowledge the computing centers and personnel of the Worldwide LHC Computing Grid for delivering so effectively the computing infrastructure essential to our analyses. Finally, we acknowledge the enduring support for the construction and operation of the LHC and the CMS detector provided by the following funding agencies: BMWF and FWF (Austria); FNRS and FWO (Belgium); CNPq, CAPES, FAPERJ, and FAPESP (Brazil); MEYS (Bulgaria); CERN; CAS, MoST, and NSFC (China); COLCIENCIAS (Colombia); MSES (Croatia); RPF (Cyprus); MoER, SF0690030s09 and ERDF (Estonia); Academy of Finland, MEC, and HIP (Finland); CEA and CNRS/IN2P3 (France); BMBF, DFG, and HGF (Germany); GSRT (Greece); OTKA and NKTH (Hungary); DAE and DST (India); IPM (Iran); SFI (Ireland); INFN (Italy); NRF and WCU (Republic of Korea); LAS (Lithuania); CINVESTAV, CONACYT, SEP, and UASLP-FAI (Mexico); MSI (New Zealand); PAEC (Pakistan); MSHE and NSC (Poland); FCT (Portugal); JINR (Armenia, Belarus, Georgia, Ukraine, Uzbekistan); MON, RosAtom, RAS and RFBR (Russia); MSTD (Serbia); SEIDI and CPAN (Spain); Swiss Funding Agencies (Switzerland); NSC (Taipei); ThEPCenter, IPST and NSTDA (Thailand); TUBITAK and TAEK (Turkey); NASU (Ukraine); STFC (United Kingdom); DOE and NSF (USA).
Individuals have received support from the Marie-Curie programme and the European Research Council and EPLANET (European Union); the Leventis Foundation; the A. P. Sloan Foundation; the Alexander von Humboldt Foundation; the Belgian Federal Science Policy Office; the Fonds pour la Formation \`a la Recherche dans l'Industrie et dans l'Agriculture (FRIA-Belgium); the Agentschap voor Innovatie door Wetenschap en Technologie (IWT-Belgium); the Ministry of Education, Youth and Sports (MEYS) of Czech Republic; the Council of Science and Industrial Research, India; the Compagnia di San Paolo (Torino); the HOMING PLUS programme of Foundation for Polish Science, cofinanced by EU, Regional Development Fund; and the Thalis and Aristeia programmes cofinanced by EU-ESF and the Greek NSRF.

%% file: B2G-12-014-authorlist.tex
\textbf{Yerevan Physics Institute,  Yerevan,  Armenia}\\*[0pt]
S.~Chatrchyan, V.~Khachatryan, A.M.~Sirunyan, A.~Tumasyan
\vskip\cmsinstskip
\textbf{Institut f\"{u}r Hochenergiephysik der OeAW,  Wien,  Austria}\\*[0pt]
W.~Adam, T.~Bergauer, M.~Dragicevic, J.~Er\"{o}, C.~Fabjan\cmsAuthorMark{1}, M.~Friedl, R.~Fr\"{u}hwirth\cmsAuthorMark{1}, V.M.~Ghete, N.~H\"{o}rmann, J.~Hrubec, M.~Jeitler\cmsAuthorMark{1}, W.~Kiesenhofer, V.~Kn\"{u}nz, M.~Krammer\cmsAuthorMark{1}, I.~Kr\"{a}tschmer, D.~Liko, I.~Mikulec, D.~Rabady\cmsAuthorMark{2}, B.~Rahbaran, C.~Rohringer, H.~Rohringer, R.~Sch\"{o}fbeck, J.~Strauss, A.~Taurok, W.~Treberer-Treberspurg, W.~Waltenberger, C.-E.~Wulz\cmsAuthorMark{1}
% ...
\vskip\cmsinstskip
% ...
1:~~Also at Vienna University of Technology, Vienna, Austria\\
2:~~Also at CERN, European Organization for Nuclear Research, Geneva, Switzerland\\
% ...
